\def\ra{\rangle}
\def\la{\langle}
\def\be{\begin{equation}}
\def\ee{\end{equation}}
\def\ba{\begin{array}}
\def\ea{\end{array}}

\def\Cb{{\Bbb C}}

\documentclass[preprint,showpacs,amsmath]{revtex4}
\usepackage{amssymb}
\def\qed{\leavevmode\unskip\penalty9999 \hbox{}\nobreak\hfill
     \quad\hbox{\leavevmode  \hbox to.77778em{%
               \hfil\vrule   \vbox to.675em%
               {\hrule width.6em\vfil\hrule}\vrule\hfil}}
     \par\vskip3pt}

\newtheorem{lemma}{Lemma}
\input amssym.def
\begin{document}

\begin{center}
{\bf A Note on State Decomposition Independent Local Invariants}
\end{center}

\begin{center}
{Ting-Gui Zhang$^{1\dag}$}, {Naihuan Jing$^{2, 3}$}, {Xianqing
Li-Jost$^{1}$}, {Ming-Jing Zhao$^{1}$}, {Shao-Ming Fei$^{1,4}$}
\vspace{2ex}

\begin{minipage}{6in}

\small $~^{1}$ {Max-Planck-Institute for Mathematics in the
Sciences, 04103 Leipzig, Germany}

{\small $~^{2}$ Department of Mathematics, North Carolina State
University, Raleigh, NC 27695, USA}

{\small $~^{3}$ School of Sciences, South China University of
Technology, Guangzhou 510640, China}

{\small $~^{4}$ School of Mathematical Sciences, Capital Normal
University, Beijing 100048, China}

 {\small $~^{\dag}$E-mail: tinggui333@163.com}

\end{minipage}
\end{center}

{\bf Abstract.} We derive a set of invariants under local unitary
transformations for arbitrary dimensional quantum systems. These
invariants are given by hyperdeterminants and independent from the
detailed pure state decompositions of a given quantum state. They
also give rise to necessary conditions for the equivalence of
quantum states under local unitary transformations.


\section{Introduction}

Invariants under local unitary transformations are tightly related
to the discussions on nonlocality - a fundamental phenomenon in
quantum mechanics, to the quantum entanglement and classification of
quantum states under local transformations. In recent years many
approaches have been presented to construct invariants of local
unitary transformations. One method is developed in terms of
polynomial invariants Refs. \cite{Rains,Grassl}, which allows in
principle to compute all the invariants of local unitary
transformations, though it is not easy to perform operationally. In
Ref. \cite{makhlin}, a complete set of 18 polynomial invariants is
presented for the locally unitary equivalence of two qubit-mixed
states. Partial results have been obtained for three qubits states
\cite{linden}, tripartite pure and mixed states \cite{SCFW}, and
some generic mixed states \cite{SFG, SFW, SFY}. Recently the local
unitary equivalence problem for multiqubit \cite{mqubit} and general
multipartite \cite{B. Liu} pure states has been solved.

However, generally one still has no operational ways to judge the
equivalence of two arbitrary dimensional bipartite or multipartite
mixed states under local unitary transformations. An effective way
to deal with the local equivalence of quantum states is to find the
complete set of invariants under local unitary transformations.
Nevertheless usually these invariants depend on the detailed
expressions of pure state decompositions of a state. For a given
state, such pure state decompositions are infinitely many.
Particularly when the density matrices are degenerate, the problem
becomes more complicated. Since in this case even the eigenvector
decompositions of a given state are not unique.

In this note, we give a way of constructing invariants under local
unitary transformations such that the invariants obtained in this
way are independent from the detailed pure state decompositions of a
given state. They give rise to operational necessary conditions for
the equivalence of quantum states under local unitary
transformations. We show that the hyperdeterminants, the generalized
determinant for higher dimensional matrices \cite{gel}, can be used
to construct such invariants. The hyperdeterminant is in fact
closely related to the entanglement measure like concurrence
\cite{Hill,wott,uhlm,rungta,ass} and 3-tangle \cite{coff}. It has
also been used in classification of multipartite pure state
\cite{miy,Luq,vie}. By employing hyperdeterminants, we construct
some trace invariants that are independent of the detailed pure
state decompositions of a given state. These trace invariants are a
priori invariant under local unitary transformations.

\section{State decomposition independent local invariant}

Let $H_1$ and $H_2$ be $n$ and $m$-dimensional complex Hilbert
spaces, with $\{\vert i\rangle\}_{i=1}^n$ and $\{\vert
j\rangle\}_{j=1}^m$ the orthonormal basis of spaces $H_1$ and $H_2$
respectively. Let $\rho$ be an arbitrary mixed state defined on
$H_1\otimes H_2$,
\begin{eqnarray}\label{general decomposition-1}
\rho=\sum_{i=1}^I p_i|v_i\ra\la v_i|,
\end{eqnarray}
where $|v_i\ra$ is a normalized bipartite pure state of the form:
$$
|v_i\ra=\sum_{k,l=1}^{n,m}a_{kl}^{(i)}|kl\ra,\ \
\sum_{k,l=1}^{n,m}a_{kl}^{(i)}a_{kl}^{(i)\ast}=1,\ \ a_{kl}^{(i)}\in
\Cb,
$$
where $\ast$ denotes complex conjugation. Denote $A_i$ the matrix
with entries given by the coefficients of the vector
$\sqrt{p_{i}}|v_i\ra$, i.e. $(A_i)_{kl}=(\sqrt{p_{i}}a_{kl}^{(i)})$
for all $i=1,\cdots,I$. Define $I\times I$ matrix $\Omega$ such that
$(\Omega)_{ij}=tr(A_iA_{j}^{\dag}), ~~~ i,j=1,\cdots,I,$ \  where
$\dag$ stands for transpose and complex conjugation.

The pure state decomposition (\ref{general decomposition-1}) of a
given mixed state $\rho$ is not unique. For another decomposition:
\begin{eqnarray}\label{general decomposition-2}
\rho=\sum_{i=1}^I q_i|\psi_i\ra\la\psi_i|,
\end{eqnarray}
with
$$
|\psi_i\rangle=\sum_{k,l=1}^{n,m}b_{kl}^{(i)}|kl\ra,\ \
\sum_{k,l=1}^{n,m}b_{kl}^{(i)}b_{kl}^{(i)\ast}=1,\ \
b_{kl}^{(i)}\in\Cb,
$$
one similarly has matrices $B_i$ with entries
$(B_i)_{kl}=(\sqrt{q_{i}}b_{kl}^{(i)})$, $i=1,\cdots,I$, and
$I\times I$ matrix $\Omega^\prime $ with entries
$$
(\Omega^\prime )_{ij}=tr(B_iB_{j}^{\dag}),~~~ \ i,j=1,\cdots,I.
$$

A quantity $F(\rho)$ is said to be invariant under local unitary
transformations if $F(\rho)=F((u_1\otimes u_2)\rho (u_1\otimes
u_2)^\dag)$ for any unitary operators $u_1\in SU(n)$ and $u_2\in
SU(m)$. Generally $F(\rho)$ may depend on the detailed pure state
decomposition. We investigate invariants $F(\rho)$ that are
independent on the detailed decompositions of $\rho$. That is,
expression in Eq. (\ref{general decomposition-1}) and expression in
Eq. (\ref{general decomposition-2}) give the same value of $F(\rho)$
for a given state $\rho$. These kind of invariants are of special
significance in determining the equivalence of two density matrices
under local unitary transformations.

Two density matrices $\rho$ and $\tilde{\rho}$ are said to be
equivalent under local unitary transformations if there exist
unitary operators $u_1$ (resp. $u_2$) on the first (resp. second)
space of $H_1\otimes H_2$ such that \be\label{lu} \tilde{\rho}=
(u_1\otimes u_2)\rho(u_1\otimes u_2)^\dag. \ee

A necessary condition that (\ref{lu}) holds is that the local
invariants have the same values $F(\rho)=F(\tilde{\rho})$. Therefore
if the expression of the invariants $F(\rho)$ do not depend on the
detailed pure state decomposition, one can easily compare the values
of the invariants between $F(\rho)$ and $F(\tilde{\rho})$. Otherwise
one has to verify $F(\rho)=F(\tilde{\rho})$ by surveying all the
possible pure state decompositions of $\rho$ and $\tilde{\rho}$. In
particular, when $\rho$ is degenerate, even the eigenvector
decomposition is not unique, which usually gives rise to the main
problem in finding an operational criterion for local equivalence of
quantum states. In fact, we have presented a complete set of
invariants in \cite{ZGFJL}. However, these invariants depend on the
eigenvectors of a state $\rho$. When the state is degenerate, this
set of invariants is no longer efficient as criterion of local
equivalence.

We set out to discuss how to find parametrization independent local
unitary invariants. First of all we give an elementary result  that
the determinant can be used to give invariants that are independent
from the choice of the pure state decomposition.

\noindent{ \bf Theorem 1:} The coefficients $F_i(\Omega)$,
$i=1,2,...,I$, of the characteristic polynomials of the matrix
$\Omega$,
\begin{eqnarray}\label{thm}
\det(\Omega-\lambda\,E)= \lambda^I + \lambda^{I-1} F_1(\Omega) +
\cdots + \lambda F_{I-1}(\Omega)+ F_{I}(\Omega) =
\Sigma_{i=0}^{I}\lambda^{I-i}F_{i}(\Omega),
\end{eqnarray}
where $E$ is the $I \times I$ unit matrix, $F_{0}(\Omega)=1$, $\det$
denotes the determinant, have the following properties:

(i) $F_{i}(\Omega)$ are independent of the pure state decompositions
of $\rho$;

(ii) $F_{i}(\Omega)$ are invariant under local unitary
transformations, $i=1,\cdots, I$.

\noindent{ \bf Proof:} (i) If Eq. (\ref{general decomposition-1})
and Eq. (\ref{general decomposition-2}) are two different
representations of a given mixed state $\rho$, we have
$B_{i}=\Sigma_{j} U_{ij}A_{j}$ for some unitary operator
$U$\cite{nie}. Consequently,
$$
\ba{rcl} \Omega_{ij}^{\prime}&=& \displaystyle tr(B_{i}B_{j}^{\dag})
= tr\left[\sum_{k,l}U_{ik}A_{k}U_{jl}^\ast A_{l}^{\dag}\right]\\[5mm]
&=&\displaystyle \sum_{k,l}U_{ik}U_{jl}^\ast\, tr(A_{k}A_{l}^{\dag})
=\sum_{k,l}U_{ik}U_{jl}^\ast \Omega_{kl} =(U\Omega U^{\dag})_{ij},
\ea
$$
i.e. $\Omega^\prime =U\Omega U^{\dag}$. Therefore
$\det(\Omega^\prime -\lambda\,E)=\det( U\Omega
U^{\dag}-\lambda\,E)=\det(\Omega-\lambda\,E)$. Thus the matrices
$\Omega$ and $\Omega^\prime $ have the same characteristic
polynomials. Namely $F_{i}(\Omega)=F_{i}(\Omega^\prime )$. Therefore
$F_{i}(\Omega)$ are invariants under the pure state decomposition
transformations.

(ii) Let $P\otimes Q\in SU(n)\otimes SU(m)$. Under the local unitary
transformations one has
$$\tilde{\rho}=(P\otimes Q)\rho (P\otimes
Q)^{\dag}=\sum_{i=1}^{I}p_i(P\otimes Q)|v_i\ra\la v_i| (P\otimes
Q)^{\dag}=\sum_{i=1}^{I}p_i|w_i\ra\la w_i|,$$ with
$$|w_i\ra=P\otimes Q|v_i\ra=\sum_{k,l=1}^{n,m}a_{kl}^{(i)\prime
}|kl\ra,\ \ \sum_{k,l=1}^{n,m}a_{kl}^{(i)\prime }a_{kl}^{{(i)\prime
}\ast}=1,\ \ a_{kl}^{(i)\prime }\in \Cb.$$ Denote
$(A_i^{\prime})_{kl}=\sqrt{p_i}a_{kl}^{(i)\prime}$. We have \be
A_{i}^{\prime}=PA_iQ^{T}. \ee Therefore
$tr(A_iA_{j}^{\dag})=tr(A_i^{\prime}A_{j}^{\prime \dag})$ and
$\Omega(\rho)=\Omega(\tilde{\rho})$. Hence
$F_{i}(\Omega(\rho))=F_{i}(\Omega(\tilde{\rho}))$, and
$F_{i}(\Omega)$, $i=1,\cdots, I$, are invariant under local unitary
transformations. \qed

In particular, the invariants $F_1= \sum tr(\sum_i A_iA_{i}^{\dag})$
and $F_I = \det (\Omega)$. For the case $I=2$, one has
$$\Omega=\left(
\begin{array}{cc}
tr(A_1A_1^{\dag}) & tr(A_1A_2^{\dag}) \\
tr(A_2A_1^{\dag}) & tr(A_2A_2^{\dag}) \\
\end{array}
\right)$$ and $F_1=tr(A_1A_1^{\dag})+tr(A_2A_2^{\dag})$,
$F_2=tr(A_1A_1^{\dag})tr(A_2A_2^{\dag})-tr(A_1A_2^{\dag})tr(A_2A_1^{\dag})$.

\noindent{ \bf Remark:} The number of local invariants $F_i$ is
uniquely determined by the rank $r$ of the mixed state $\rho$, i.e.
$I=r$. Therefore we only need to calculate the invariants
corresponding to the eigenvector decomposition. Because for
arbitrary pure state decomposition $\rho=\Sigma_{j=1}^{J}
q_j|\psi_j\rangle\langle\psi_j|$ with $J>r$, the above determinant
is the same as that of the eigenvector decomposition of
$\rho=\Sigma_{i}^r p_i|\phi_i\rangle\langle\phi_i|$ after adding
$J-r$ zero vectors. The determinant $\det(\Omega^\prime
-\lambda\,E)$ of the eigenvector decomposition of $\rho$ after
adding $J-r$ zero vectors and $\det(\Omega-\lambda\,E)$ of
$\rho=\Sigma_{j=1}^{r} q_j|\psi_j\rangle\langle\psi_j|$ without
$J-r$ zero vectors have the relation: $\det(\Omega^\prime
-\lambda\,E)=\lambda^{J-r}\det(\Omega-\lambda\,E)$. This means that
the number of independent local invariants given by (\ref{thm}) does
not depend on the number of pure states in the ensemble of a given
$\rho$. Therefore if two mixed states $\rho$ and $\tilde{\rho}$ have
different ranks, they are not local unitary equivalent. If their
ranks are the same, one only needs to calculate the corresponding
invariants with respect to the same numbers $I$ of pure states in
the pure state decompositions.

In fact for a quantum state $\rho$ in eigenvector decomposition
$\rho=\sum_i \lambda_i |\psi_i\rangle\langle\psi_i|$, the
corresponding matrix $\Omega$ is a diagonal one with $\rho$'s
eigenvalues $ \lambda_i$ as the diagonal entries. In this case the
local invariants from Theorem 1 are just the coefficients of the
characteristic polynomial of the quantum state $\rho$. Theorem 1
shows that these coefficients are local invariants and independent
from the detailed pure state decompositions. But the easy approach
employed in Theorem 1 can be generalized to construct more local
invariants that are independent of the detailed pure state
decompositions by using hyperdeterminant \cite{gel}.

In order to derive more parametrization independent quantities we
consider the multilinear form $f_A: \underbrace{V\otimes \cdots
\otimes V}_\text{$2s$}\mapsto \mathbb C$ given by
\begin{equation}\label{eq:form}
f_A(e_{i_1}, \cdots, e_{i_s}, e_{j_1}, \cdots,
e_{j_s})=tr(A_{i_1}A_{j_1}^{\dagger}\cdots
A_{i_s}A_{j_s}^{\dagger}),
\end{equation}
where $e_i$ ($1\leq i\leq I$) are standard basis elements in
$V={\mathbb C}^I$. The multilinear form $f$ can also be written as a
tensor in $V^*\otimes\cdots \otimes V^*$:
\begin{equation}\label{eq:form2}
f_A=\sum_{\underline{i},
\underline{j}}tr(A_{i_1}A_{j_1}^{\dagger}\cdots
A_{i_s}A_{j_s}^{\dagger})e_{i_1}^*\otimes\cdots\otimes
e_{i_s}^*\otimes e_{j_1}^*\otimes\cdots\otimes e_{j_s}^*,
\end{equation}
where $e_i^*$ are standard $1$-forms on ${\mathbb C}^I$ such that
$e_i^*(e_j)=\delta_{ij}$, and $\underline{i}=(i_1, \cdots, i_s),
\underline{j}=(j_1, \cdots, j_s), 1\leq i_p, j_p\leq I$. In general
we call the $2s$-dimensional matrix or hypermatrix
$A=(A_{\underline{i}\underline{j}})=(tr(A_{i_1}A_{j_1}^{\dagger}\cdots
A_{i_s}A_{j_s}^{\dagger}))$ formed by the coefficients of
(\ref{eq:form2}) the hypermatrix of the multilinear form $f_A$
relative to
the standard basis.  

The Cayley hyperdeterminant Det($A$) \cite{gel} is defined to be the
resultant of the multilinear form $f_A$, that is, Det(A) is the
normalized integral equation of the hyperplane given by the
multilinear form $f_A$. It is known that \cite{gel} the
hyperdeterminant exists for a given format and is unique up to a
scalar factor, if and only if the largest number in the format is
less than or equal to the sum of the other numbers in the format.
Hyperdeterminants enjoy many of the properties of determinants. One
of the most familiar properties of determinants, the multiplication
rule $det(AB) = det(A) det(B)$, can be generalized to the situation
of hyperdeterminants as follows. Given a multilinear form
$f(x^{(1)}, ..., x^{(r)})$ and suppose that a linear transformation
acting on one of its components using an $n\times n$ matrix B, $y_r
= B x_r$. Then
\begin{equation}\label{eq:det-action}
Det(f.B) = Det(f) det(B)^{N/n},
\end{equation}
where $N$ is the degree of the hyperdeterminant. Therefore we have
the following result.

\begin{lemma} The hyperdeterminant of format $(k_1,\ldots,k_r)$ is an invariant under
the action of the group $SL(k_1) \otimes \cdots \otimes SL(k_r)$,
and subsequently also invariant under $SU(k_1) \otimes \cdots
\otimes SU(k_r)$.
\end{lemma}
\noindent{ \bf Proof:} For $(A, B, \cdots, C)\in SL(k_1) \otimes
\cdots \otimes SL(k_r)$, it follows from Eq. (\ref{eq:det-action})
that

\begin{align}\label{eq:det-eq}
Det((A_{(1)}\cdot B_{(2)}\cdot\cdots C_{(r)}\cdot)f) = Det(f)
det(A)^{N/k_1}det(B)^{N/k_2}\cdots det(C)^{N/k_r} =Det(f).
\end{align}

The three-dimensional hyperdeterminant of the format $2\times
2\times 2$ is known as the Cayley's hyperdeterminant \cite{Ca}. In
this case the hyperdeterminant of a hypermatrix $A$ with components
$a_{ijk}$, $i,j,k \in \{0, 1\}$, is given by
\begin{eqnarray}
Det(A) &=& a_{000}^2a_{111}^2 + a_{001}^2a_{110}^2 +
a_{010}^2a_{101}^2 + a_{100}^2a_{011}^2 -
2a_{000}a_{001}a_{110}a_{111}\\\nonumber
&&-2a_{000}a_{010}a_{101}a_{111}-2a_{000}a_{011}a_{100}a_{111} -
2a_{001}a_{010}a_{101}a_{110}
\\\nonumber &&-2a_{001}a_{011}a_{110}a_{100}- 2a_{010}a_{011}a_{101}a_{100} +
4a_{000}a_{011}a_{101}a_{110}\\\nonumber && +
4a_{001}a_{010}a_{100}a_{111}.
\end{eqnarray}
This hyperdeterminant can be written in a more compact form by using
the Einstein convention and the Levi-Civita symbol
$\varepsilon^{ij}$, with $\varepsilon^{00} =\varepsilon^{11} = 0,
\varepsilon^{01} = -\varepsilon^{10} = 1$; and $b_{kn} =
(1/2)\varepsilon^{il}\varepsilon^{jm}a_{ijk}a_{lmn}$, $ Det(A)
=(1/2)\varepsilon^{il}\varepsilon^{jm}b_{ij}b_{lm}$. The
four-dimensional hyperdeterminant of the format $2\times 2\times 2
\times 2$ has been given in Ref. \cite{Luq}.

For the general mixed state $\rho$ in Eq. (\ref{general
decomposition-1}), we can define a hypermatrix $\Omega_{s}$ with
entries
\begin{eqnarray}
(\Omega_{s})_{i_1i_2\cdots i_sj_1j_2\cdots j_s}
=tr(A_{i_1}A_{j_1}^{\dag}A_{i_2}A_{j_2}^{\dag}\cdots
A_{i_s}A_{j_s}^{\dag}),
\end{eqnarray}
for $i_k,j_k=1,\cdots,I$, $s \geq 1$, $0\leq i_{j}\leq k_{j}$. The
format of $\Omega_{s}$ is $I\times \cdots \times I$.

\noindent{ \bf Theorem 2:} $Det (\Omega_s-\lambda\,E)$, with
$E=(E_{i_1,i_2,
\cdots,i_s,j_1,j_2,\cdots,j_s})=(\delta_{i_1j_1}\delta_{i_2j_2}
\cdots \delta_{i_sj_s})$, is independent of the pure state
decompositions of $\rho$. It is also invariant under local unitary
transformations of $\rho$. In particular, all coefficients of
polynomial $Det (\Omega_s-\lambda\,E)$ are local invariants
independent from the pure state decompositions and are invariance
under local unitary transformations.

\noindent{ \bf Proof:} We first show that it is independent from the
pure state decomposition of $\rho$. Let Eq. (\ref{general
decomposition-1}) and Eq. (\ref{general decomposition-2}) be two
different representations of a given mixed state $\rho$. We have
\begin{eqnarray}
(\Omega_s^\prime )_{i_1i_2\cdots i_sj_1j_2\cdots
j_s}&=&tr(B_{i_1}B_{j_1}^{\dag}B_{i_2}B_{j_2}^{\dag}\cdots
 B_{i_s}B_{j_s}^{\dag})\\\nonumber
&=&tr\left[\Sigma_{i_1^\prime j_1^\prime, \cdots, i_s^\prime
 j_s^\prime} U_{i_1i_1^\prime}A_{i_1^{\prime}}U_{j_1j_1^\prime}^\ast
A_{j_1^{\prime}}^{\dag}\cdots
U_{i_si_s^\prime}A_{i_s^{\prime}}U_{j_sj_s^\prime}^\ast
A_{j_s^{\prime}}^{\dag}\right]\\\nonumber &=&((U \otimes U \otimes
\cdots \otimes U) (\Omega_{s})(U^{\dag} \otimes U^{\dag} \otimes
\cdots \otimes U^{\dag}))_{i_1i_2\cdots i_sj_1j_2\cdots j_s}.
\end{eqnarray}
Therefore $\Omega^\prime _s=(U \otimes U \otimes \cdots \otimes U)
\Omega_{s} (U^{\dag} \otimes U^{\dag} \otimes \cdots \otimes
U^{\dag})$. Using the action, the associated multilinear form
$f_{\omega}$ is acted upon by $U \otimes U \otimes \cdots \otimes U$
and $U^{\dag} \otimes U^{\dag} \otimes \cdots \otimes U^{\dag}$ as
follows:
\begin{equation*}
(U_{(1)}\cdot \cdots U_{(s)}\cdot U^{\ast}_{(1)}\cdot \cdots
U^{\ast}_{(s)}\cdot) f_{\omega}
\end{equation*}
Using the formula under the action (\ref{eq:det-eq}) we get $Det
(\Omega^\prime _s-\lambda\,E)=Det (\Omega_s-\lambda\,E)$, and thus
$Det (\Omega_s-E\,\lambda)$ does not depend on the detailed pure
state decompositions of a given $\rho$. Note that in general we
don't know the exact formula for the hyperdeterminant, but we can
still derive its invariance abstractly.

On the other hand, under local unitary transformations
$\tilde{\rho}=(P\otimes Q)\rho(P\otimes Q)^\dag$ for some local
unitary operators $P\otimes Q\in SU(n)\otimes SU(m)$, similar to the
proof of the second part of the Theorem 1 and using Lemma
\ref{eq:det-action} in \cite{gel}, it is easy to get
$\Omega_s=\Omega^\prime _s$. Therefore $Det(\Omega_s-\lambda\,E)$ is
invariant under local unitary transformations. Moreover,
following\cite{Luq}, the invariant polynomials are invariance under
local unitary transformations. \qed

As the application of our theorems we now give two interesting
examples.

Example 1: Consider two mixed states $\rho_1=diag\{1/2,1/2,0,0\}$
and $\rho_2=diag\{1/2,0,1/2,0\}$. $\rho_1$ has a pure state
decomposition with
$$
A_0=\left(
\begin{array}{cc}
\frac{1}{\sqrt{2}} & 0 \\
0 & 0 \\
\end{array}
\right),~~~~ A_1=\left(
\begin{array}{cc}
 0 & \frac{1}{\sqrt{2}} \\
  0 & 0
   \end{array}
    \right).
$$
While $\rho_2$ has a pure state decomposition with
$$
B_0=\left(
 \begin{array}{cc}
 \frac{1}{\sqrt{2}} & 0 \\
  0 & 0 \\
   \end{array}
    \right),~~~~
     B_1=\left(
     \begin{array}{cc}
      0 & 0 \\
       \frac{1}{\sqrt{2}} & 0
        \end{array}
        \right).
$$
We have the corresponding matrices
$(\Omega(\rho_1))_{i,j}=tr(A_iA_{j}^{\dag})$ and $(\Omega
(\rho_2))_{i,j}=tr(B_iB_{j}^{\dag})$, $i,j=0,1$. From Theorem 1 one
can find that these two states have the same values of the
invariants in Eq. (\ref{thm}), $F_i(\Omega(\rho_1))
=F_i(\Omega(\rho_2))$.

We now consider further the four-dimensional hyperdeterminant of the
format $2\times 2\times 2 \times 2$ \cite{Luq}. Let
$(\Omega(\rho_1))_{ijkl}=tr(A_iA_{j}^{\dag}A_kA_{l}^{\dag})\equiv
a_r$, $r=0,\cdots,15$, where $r=8i+4j+2k+l$. From Ref. \cite{Luq},
one invariant with degree $4$ is given by
$$
N(\rho_1)=det\left(
\begin{array}{cccc}
a_{0} & a_{1} & a_{8} & a_{9} \\
 a_{2} & a_{3} & a_{10} & a_{11} \\
  a_{4} & a_{5} & a_{12} & a_{13} \\
   a_{6} & a_{7} & a_{14} & a_{15}
    \end{array}
\right)=\frac{1}{256}.
$$
However for $\rho_{2}$ we have $N(\rho_{2})=0$. Therefore $\rho_1 $
and $\rho_{2}$ are not equivalent under local unitary
transformations.

In Ref. \cite{che}, the Ky Fan norm of the realignment matrix of the
quantum states $\cal{N}(\rho)$ is proved to be invariant under local
unitary operations. By calculation we find
${\cal{N}}(\rho_1)={\cal{N}}(\rho_2)=\frac{1}{\sqrt{2}}$. This means
the Ky Fan norm of the realignment matrix can not recognize that
$\rho_1$ and $\rho_2$ are not equivalent under local unitary
transformations. Therefore Theorem 2 has its superiority over it
with respect to these two states.

Example 2: Let two mixed states $\sigma_1=\left(
\begin{array}{cccc}
\frac{1}{3} & 0 & 0 & 0 \\
 0 & \frac{1}{3} & \frac{1}{3} & 0 \\
 0 & \frac{1}{3} & \frac{1}{3} & 0 \\
 0 & 0 & 0 & 0
    \end{array}
\right)$ and $\sigma_2=diag\{2/3,0,0,1/3\}$. Then $\sigma_1$ has a
pure state decomposition with
$$
C_0=\left(
\begin{array}{cc}
\frac{1}{\sqrt{3}} & 0 \\
0 & 0 \\
\end{array}
\right),~~~~ C_1=\left(
\begin{array}{cc}
 0 & \frac{1}{\sqrt{3}} \\
  \frac{1}{\sqrt{3}} & 0
   \end{array}
    \right).
$$
While $\sigma_2$ has a pure state decomposition with
$$
D_0=\left(
 \begin{array}{cc}
 \frac{\sqrt{2}}{\sqrt{3}} & 0 \\
  0 & 0 \\
   \end{array}
    \right),~~~~
     D_1=\left(
     \begin{array}{cc}
      0 & 0 \\
       0 & \frac{1}{\sqrt{3}}
        \end{array}
        \right).
$$
We have the corresponding matrices
$(\Omega(\sigma_1))_{i,j}=tr(C_iC_{j}^{\dag})$ and $(\Omega
(\sigma_2))_{i,j}=tr(D_iD_{j}^{\dag})$, $i,j=0,1$. From Theorem 1
one can find that these two states have the same values of the
invariants in Eq. (\ref{thm}), $F_i(\Omega(\sigma_1))
=F_i(\Omega(\sigma_2))$.

We also consider further the four-dimensional hyperdeterminant of
the format $2\times 2\times 2 \times 2$. Also let
$(\Omega(\sigma_1))_{ijkl}=tr(C_iC_{j}^{\dag}C_kC_{l}^{\dag})\equiv
a_s$, $r=0,\cdots,15$, where $s=8i+4j+2k+l$. From Ref. \cite{Luq},
the another invariant with degree $4$ is given by
$$
M(\sigma_1)=det\left(
\begin{array}{cccc}
a_{0} & a_{8} & a_{2} & a_{10} \\
 a_{1} & a_{9} & a_{3} & a_{11} \\
  a_{4} & a_{12} & a_{6} & a_{14} \\
   a_{5} & a_{13} & a_{7} & a_{15}
    \end{array}
\right)=\frac{1}{6561}.
$$
However for $\sigma_{2}$ we have $M(\sigma_{2})=0$. Therefore
$\sigma_1 $ and $\sigma_{2}$ are not equivalent under local unitary
transformations. From example 2, one can see that the spectra of
their reduced one-qubit density matrices have the same value.
Therefore, only by the spectra of their reduced one-qubit density
matrices can not judge the equivalence of given states.

Our results can be generalized to multipartite case. Let
$H_1,H_2,\cdots,H_m$ be $n_1,n_2, \cdots, n_m$-dimensional complex
Hilbert spaces with $\{\vert k_1\rangle\}_{k_1=1}^{n_1}$, $\{\vert
k_2\rangle\}_{k_2=1}^{n_2}$, $\cdots $, $\{\vert
k_m\rangle\}_{k_m=1}^{n_m}$ the orthonormal basis of $H_1, H_2,
\cdots, H_m$ respectively. Let $\rho$ be an arbitrary mixed state
defined on $H_1\otimes H_2\otimes\cdots\otimes H_m$,
$\rho=\sum_{i=1}^Ip_i|v_i\ra\la v_i|$, where $|v_i\ra$ is a
multipartite pure state of the form:
$|v_i\ra=\sum_{k_1,k_2,\cdots,k_m=1}^{n_1,n_2,\cdots,n_m}a_{k_1k_2\cdots
k_m}^{(i)}|k_1k_2\cdots k_m\ra,\ \ a_{k_1k_2\cdots k_m}^{(i)}\in
\Cb$. Now we view $|v_i\ra$ as bipartite pure state under the
partition between the first $l$ subsystems and the rest, $1\leq l
<n$. Then $A_i=(\sqrt{p_{i}}a_{k_1k_2\cdots k_m}^{(i)})$ can be
regarded as the $N_1\times N_2$ matrix with $N_1=n_1\times n_2\times
\cdots\times n_l$ and $N_2=n_{l+1}\times n_{l+2}\times \cdots \times
n_m$ for all $i=1,\cdots,I$. We define the matrix $\Omega_{s}$ with
entries $(\Omega_s)_{i_1i_2\cdots i_sj_1j_2\cdots
j_s}=tr(A_{i_1}A_{j_1}^{\dag}\cdots A_{i_s}A_{j_s}^{\dag})$, for
$i_k,j_k=1,\cdots,I$, $s \geq 1$. Then we have that
$Det(\Omega_s-\lambda\,E)$ does not depend on the pure states
decompositions and is invariant under local unitary transformations.

\section{Conclusion}

We have investigated the invariants under local unitary
transformations for arbitrary dimensional quantum systems. These
invariants are independent of the detailed pure state decompositions
of a given state. They give rise to the necessary conditions for the
equivalence of quantum states under local unitary transformations.
These invariants may be also used in characterizing quantum
correlations such as quantum entanglement \cite{quant-ent} and
quantum discord \cite{quant-disc}, since all these quantities are at
least the invariants under local unitary transformations and are
independent of the pure state decompositions.

Acknowledgments: This work is supported by the Simons Foundation
under  No. 198129 and the National Natural Science Foundation of
China under No. 11271138, No. 11275131.

\end{document}